# Lifshitz transition mediated electronic transport anomaly in bulk ZrTe$_5$

Hang Chi[1†], Cheng Zhang[1,2†], Genda Gu[1], Dmitri E. Kharzeev[3,4], Xi Dai[5,6] and Qiang Li[1*]

[1]*Condensed Matter Physics and Materials Science Department,*
*Brookhaven National Laboratory, Upton, New York 11973, USA*
[2]*Materials Science and Engineering Department, Stony Brook University, Stony Brook, NY 11794, USA*
[3]*Department of Physics, Brookhaven National Laboratory, Upton, New York 11973, USA*
[4]*Department of Physics and Astronomy, Stony Brook University, New York 11794, USA*
[5]*Beijing National Laboratory for Condensed Matter Physics,*
*Institute of Physics, Chinese Academy of Sciences, Beijing 100190, China*
[6]*Collaborative Innovation Center of Quantum Matter, Beijing 100871, China*

[†]These authors contributed equally to this work.

[*]Correspondence should be addressed to Q.L. (email: qiangli@bnl.gov).

**Abstract**

Zirconium pentatelluride ZrTe$_5$, a fascinating topological material platform, hosts exotic chiral fermions in its highly anisotropic three-dimensional Dirac band and holds great promise advancing the next-generation information technology. However, the origin underlying its anomalous resistivity peak has been under debate for decades. Here we provide transport evidence substantiating the anomaly to be a direct manifestation of a Lifshitz transition in the Dirac band with an ultrahigh carrier mobility exceeding $3\times10^5$ cm$^2$ V$^{-1}$ s$^{-1}$. We demonstrate that the Lifshitz transition is readily controllable by means of carrier doping, which sets the anomaly peak temperature $T_\mathrm{p}$. $T_\mathrm{p}$ is found to scale approximately as $n_\mathrm{H}^{0.27}$, where the Hall carrier concentration $n_\mathrm{H}$ is linked with the Fermi level by $\varepsilon_\mathrm{F} \propto n_\mathrm{H}^{1/3}$ in a linearly dispersed Dirac band. This relation indicates $T_\mathrm{p}$ monotonically increases with $\varepsilon_\mathrm{F}$, which serves as an effective knob for fine tuning transport properties in pentatelluride-based Dirac semimetals.
1. **Introduction**

Three-dimensional (3D) Dirac semimetals (DSMs) [1-11] and Weyl semimetals (WSMs) [12-16] have recently attracted tremendous attention, enabling investigations of quantum dynamics of relativistic field theory in condensed matter experiments. The relativistic theory of charged chiral fermions (massless spin 1/2 particles with a definite projection of spin on momentum) in 3D possesses the chiral anomaly – non-conservation of chiral charge induced by external gauge fields with nontrivial topology, e.g., by parallel electric and magnetic fields (**E** ∥ **B**). The chiral quasiparticles in DSMs and WSMs have opened unprecedented opportunities to study the effects of the chiral anomaly. Of particular importance is the chiral magnetic effect (CME) – observation of a chirality-imbalance-induced electric current in the presence of an external

Page 2 of 17

magnetic field, see figure 1 [17-23]. The signature of the CME in Dirac systems with **E** ∥ **B** is a positive contribution to the electrical conductivity that has a quadratic magnetic field dependence. This is because the CME current is proportional to the product of the chirality imbalance and the magnetic field, where the chirality imbalance in Dirac systems is dynamically generated through the quantum anomaly with a rate that is proportional to the scalar product of electric and magnetic fields. The longitudinal magnetoresistance (LMR) thus becomes negative [9, 24, 25]. These extraordinary relativistic quasiparticles bode well for both fundamental exploration and practical applications [26-29].

The observation of CME in ZrTe$_5$ [9] kicked off intensive transport studies in DSM (such as Na$_3$Bi [24], Cd$_3$As$_2$ [25], ZrTe$_5$ [11, 30-37] and HfTe$_5$ [38-41]) and WSM (e.g., TaAs [42], NbAs [43], NbP [44], and TaP [45]) materials. The readily accessible quantum limit in ZrTe$_5$ (only several Tesla [10, 37], instead of above 40 T for Cd$_3$As$_2$ [46]), along with its chemical stability, advocates unique strategic advantages of exploring ZrTe$_5$-based materials as a unique topological platform [8, 47]. ZrTe$_5$ crystallizes in a layered orthorhombic *Cmcm* ($D_{2h}^{17}$, No. 63) structure, as shown in the inset of figure 2(a). Density functional theoretical (DFT) results [8, 11, 37] have shown that the electronic structure of ZrTe$_5$ is sensitive to external parameters such as pressure, temperature, and stress field introduced by chemical dopant. For example, movement of the chemical potential $\mu(T)$ in ZrTe$_5$ upon varying temperature *T* has been detected in several recent angle-resolved photoemission spectroscopy (ARPES) experiments [48-52]. Note the exact temperature dependence of $\mu(T)$, i.e., shifting up [49, 50] or down [48, 51] when *T* increases, is still much under debate. Detailed transport studies, especially on bulk samples with intrinsically low background carrier concentrations, are very desirable. Such transport experiments may provide insights towards understanding the Lifshitz transition and the Van Hove singularity as observed in previous ARPES [50, 51] and optical studies [53].

2. **Experimental**

Single crystalline ZrTe$_5$ samples were prepared via a Te-flux method. High purity Zr and Te elemental mixture Zr$_{0.0025}$Te$_{0.9975}$ were sealed under vacuum in a double-walled quartz ampule and first melted at 900 °C in a box furnace and fully rocked to achieve homogeneity for 72 hours. The melt was followed by slow cooling and rapid heating treatment between 445 °C and 505 °C for 21 days, in order to re-melt crystals with small sizes. The resultant single crystals were typically about 0.1 × 0.3 × 20 mm$^3$. Crystals were chemically and structurally analyzed by powder X-ray diffraction (XRD), scanning electron microscopy (SEM) with energy dispersive x-ray spectroscopy (EDS), and transmission electron microscopy (TEM), as described in Ref. [9].

The field dependent resistivity tensor $\rho_{xx}$ and $\rho_{yx}$ of ZrTe$_5$, in the temperature range of 5 – 300 K, were measured using the Hall-bar configurations in a Quantum Design Physical Property Measurement System (PPMS) equipped with a 9 Tesla superconducting magnet. The zero-field Seebeck coefficient was measured via the standard four-probe method in the Thermal Transport Option (TTO) in PPMS. Measurements were performed with the electric current and/or thermal gradient along the crystallographic *a*-axis.

3. **Results and Discussion**

As shown in figure 2(a), the electrical resistivity $\rho_{xx}$ in our bulk single crystal ZrTe$_5$ is about 0.8 mΩ cm at 5 K, and displays a metallic temperature dependence dominated by electrons, as evidenced by the negative Hall coefficient $R_H$ [figure 2(b), obtained from the zero-field *B*- derivative of the Hall resistivity $\rho_{yx}$ in figure 3] and Seebeck coefficient $S$ [figure 2(c)]. The derived Hall carrier concentration $n_H$ [$\equiv 1/(eR_H)$, negative (positive) for electrons (holes), with $e$ being the elementary charge] = $-2.6\times10^{16}$ cm$^{-3}$ is among the lowest ever achieved in ZrTe$_5$ reported so far (supplementary table 1), leading to a much lowered Fermi level $\varepsilon_F$ [9]. This is likely due to the fact that our single crystals grown in self flux are free from contamination of foreign species such as the iodine transport agent in other vapor transport crystal growth method, where iodine may act as electron dopants in ZrTe$_5$. Upon increasing temperature, the sign of both $R_H$ and $S$ switches from negative to positive around the same temperature $T_p$ = 60 K, where $\rho_{xx}$ exhibits a peak and begins to decrease. With further increasing temperature, yet another turn in $\rho_{xx}$ becomes evident, where $\rho_{xx}$ starts to increase again above approximately 200 K. The peak behavior at $T_p$ in $\rho_{xx}$ was considered as an "anomaly" [54], whose origin has been under debate [55-60].

As shown in figures 3(a) and 3(b), $\rho_{yx}(B)$ possess remarkable temperature and field dependence. At 5 K, the negative slope of the low-field $\rho_{yx}(B)$ indicates the dominance of electrons, consistent with the negative $S$ in figure 1(c). The nonlinearity of $\rho_{yx}(B)$, i.e., slope changes at higher $B$, indicates that additional carriers may contribute to transport. Upon warming up towards $T_p$, a similar field dependence is maintained. The increasing magnitude of $\rho_{yx}(B)$, up to 50 K, suggests the decrease of carrier concentrations, corroborating the increasing magnitude of $S$. When passing through $T_p$, the high-field $\rho_{yx}(B)$ drops dramatically in magnitude from 50 to 80 K, whose sign changes after 90 K. In the low-field region, a positive slope emerges at 70 K, signaling the sign change of the dominant carriers during the transition through $T_p$ (also consistent with the sign change of $S$). Above 100 K, the slope of $\rho_{yx}(B)$ drastically decreases and the nonlinearity fades out, restoring a nearly perfect linear $\rho_{yx}(B)$ at 300 K.

In order to decipher the complex temperature and field dependence of $\rho_{yx}(B)$, the Hall conductance $\sigma_{xy}$ [$\equiv \rho_{yx}/(\rho_{xx}\rho_{yy} + \rho_{yx}^2)$] is numerically fitted using a simplified two-band model,

$$\sigma_{xy} = \frac{n_1 e \mu_1^2 B}{1+\mu_1^2 B^2} + \frac{n_2 e \mu_2^2 B}{1+\mu_2^2 B^2}, \qquad (1)$$

where $n_1$ ($n_2$) and $\mu_1$ ($\mu_2$) are the carrier concentration [negative (positive) for electrons (holes)] and averaged in-plane mobility ($\mu_{ac} = \sqrt{\mu_a \mu_c}$) for the first (second) band with high (low) mobility. The magneto-resistivity $\rho_{xx}(B)$ has been previously measured in Ref. [9]. The in-plane anisotropy $b = \rho_{yy}/\rho_{xx} \sim \mu_a/\mu_c \sim m_c/m_a$ is taken to be a constant of 2, as experimentally determined in an L-shaped ZrTe$_5$ nano-device [36]. The model fits very well the experimental data, in all the measured temperature and magnetic field range [figure 3(c), regardless of the choice of $b$, see supplementary figures 1 and 2). The temperature dependence of the fitting parameters is plotted in figure 3(d). As schematically interpreted below in figure 4, the topology change in the Fermi pockets, due to the downward shifting of the chemical potential $\mu(T)$ [recall $\varepsilon_F \equiv \mu(T = 0\ \text{K})$], is a signature of the Lifshitz transition [61].

At the lowest temperatures, $\mu_1$ takes extremely high values exceeding $3\times10^5$ cm$^2$ V$^{-1}$ s$^{-1}$, consistent with expectations for a Dirac band. The linearly dispersed Dirac band is well documented by transport [9, 30, 31, 33, 37], ARPES [9, 48-52] and optical [10, 31, 53] studies. A secondary, less mobile (although $\mu_2$ still achieves high values of $8\times10^4$ cm$^2$ V$^{-1}$ s$^{-1}$) electron band is also at work at 5 K, which is likely originated from the off-zone-center electron pockets such as those observed in recent ARPES studies [49, 51]. The dominance of the chiral Dirac electrons over the trivial secondary electrons is evidenced by the observation of strong negative magnetoresistance at the low temperatures in electric field parallel to magnetic field due to CME [9]. Upon increasing temperature towards $T_p$, both $\mu_1$ ($\mu_2$) and $n_1$ ($n_2$) decreases in magnitude. It is consistent with a scenario where the Fermi velocity $v_\text{F}$ of the Dirac band decreases upon increasing temperature. The temperature-induced change in $v_\text{F}$ has also been observed in recent ARPES experiments [48, 51]. The broader band provides more states at lower energy, which effectively lowers $\mu(T)$. Thus, the increase of temperature leads to shrinking Fermi pockets at $\mu(T)$ for both the Dirac and the secondary bands.

When going through $T_\text{p}$, $n_2$ and $\mu_2$ in the secondary band maintain the same trend as that at lower temperatures. In the meantime, $n_1$ in the Dirac band crosses zero, signaling the Fermi pocket changes from electron-like to hole-like via passing through the Dirac point. The mobility $\mu_1$ in the lower Dirac band has a similarly impressive high value of almost $5\times10^5$ cm$^2$ V$^{-1}$ s$^{-1}$ at $T_1$ = 70 K. As temperature further increases, $\mu_1$ decreases. The continuous change of $n_1$ across zero (from *n*-type to *p*-type) and the discontinuity in $\mu_1$ at $T_\text{p}$ indicates $\mu(T)$ crossing the Dirac point. The Lifshitz transition at $T_1$ does not affect chiral Fermion transport. In fact, the CME was clearly observed above 70 K. From 70 to about 100 K [9], chiral Dirac holes take over. While both $n_1$ and $\mu_1$ maintain an uneventful temperature dependence above $T_\text{p}$, $n_2$ crosses zero at about $T_2$ = 150 K, accompanied by a dramatic increase in $\mu_2$ ($9\times10^4$ cm$^2$ V$^{-1}$ s$^{-1}$ at 150 K), which is an *n*- to *p*-type transition at $T_2$ in the secondary band. The secondary hole carriers are excited from the secondary valence band, the exact location and dispersion of which are not clear. Nevertheless, valance bands observed in recent ARPES studies mapping the whole Brillouin zone [49] seem to have the binding energy (e.g., ~ 0.6 eV along the Γ–X direction) too high to provide carriers via thermal activation. It appears that, largely due to its higher mobility, the secondary band gradually takes over and becomes more dominant in transport towards ambient temperatures. The dominance of the secondary band over the Dirac band at elevated temperature renders the signal of the CME conductivity too small to be detectable above 150 K, in comparison to the Ohmic conductivity [9]. The second upturn in $\rho_{xx}$ at about 200 K is consistent with finite Fermi surfaces. As will be discussed later, the Lifshitz-type topology change in the Dirac band is the main electronic origin behind the "puzzling" resistivity peak in $ZrTe_5$.

Figure 4 illustrates the Lifshitz transition in bulk $ZrTe_5$. The upper panel, figures (a)-(d), shows the evolution of the Dirac band and the secondary band, together with the Fermi-distribution function $f(\varepsilon)$ as a function of temperature. At zero temperature, the chemical potential $\mu(T)$ is located in the upper Dirac band which is at the zone center. Secondary bands are off-centered, where $\mu(T)$ is in the conduction band, and close to the band edge. As temperature increases, the Fermi velocity $v_F$ in the Dirac band decreases. The behavior is consistent with the reduction of the mobility $\mu_1$ observed here, which is further supported by the ARPES studies showing the broadening of the Dirac band as temperature increases [48, 51]. This increases the available states in the lower band that effectively moves the chemical potential $\mu(T)$ downwards. The hatched areas represent the excited states in the secondary bands. The reduction of carrier density and mobility upon increasing temperature is consistent with the observed metallic behavior of $\rho_{xx}(T)$ at temperatures below $T_p$.

Figures 4(e-h) depict the expected band topology at $\mu(T)$. For the Dirac band, the Fermi pocket changes from chiral electron to chiral hole at $T_1$ (nominally the same as $T_p$) where the pocket is reduced to a Dirac point. Above $T_1$, the size of the chiral hole pocket increases, this agrees with the reduction of $\rho_{xx}(T)$. For clarity, the thermal excitation in the Dirac band is not shown. The shape of the Dirac pockets is drawn to be rectangular, approximately reflecting the anisotropy of this band. We want to point out that the existence of a Dirac point is not evident from our transport studies. As such, we cannot rule out a small gap $\Delta_d$ between the upper and lower Dirac bands. Earlier calculation suggested that it is impossible to have symmetry protected

Dirac points in ZrTe$_5$, due to its specific crystal symmetry, unlike that in Na$_3$Bi and Cd$_3$As$_2$. The gap will in general induce chirality-changing transition with a rate $\Delta_d/\hbar$. The existence of intra-valley transition does not destroy the CME, although the chirality flipping transitions do reduce the magnitude of CME current. The chirality-changing rate should only be a small fraction of the quantum scattering rate $\Gamma_Q$, which can be determined from the broadening of the quasiparticle. $\Gamma_Q$ sets the absolute upper limit to the chirality-changing time, and hence the $\Delta_d$. The measurement of $\Gamma_Q$ is in the progress.

The off-centered circles at zero temperature, as shown in figure 4(e), are the electron pockets of the secondary band, although we do not know their exact shape and location. At elevated temperature, the chemical potential $\mu(T)$ falls in the gap of the secondary band. Carriers are thermally excited, shown by the hatched areas, that contribute to the transport. The dominant carriers in the secondary bands change from electrons ($\mu$ closer to the conduction band edge) to holes ($\mu$ closer to the valence band edge) near $T_2$ as temperature increases. Above $T_2$, when the secondary carriers begin to dominate the transport, the system gradually reverts back to a metallic behavior in $\rho_{xx}(T)$. This behavior was observed up to 400 K, as shown in figure 1(a) of Ref. [9].

In order to testify the validity and universality of the observed Lifshitz transition, literature results [30, 32, 33, 57, 59] on $\rho_{xx}$ and Seebeck coefficient $S$ are compared against our data in figures 5(a) and 5(b), respectively. The numerical values of $n_H$ are tabulated in supplementary table 1. Likely originated from the temperature dependence of the lattice parameter, the broadening of the Dirac band (decreasing $v_F$ with increasing $T$) should be an intrinsic property of ZrTe$_5$, regardless of the level of (un)intentional doping which sets the position of $\varepsilon_F$ at 0 K. Indeed, $T_p$ coincides with the transition temperature where $S$ changes sign, in nominally pure

ZrTe$_5$ compounds [59]. Intentional *p*-doping (by substituting Te with Sb [59]) can effectively "switch off" the Lifshitz transition, since with $\mu(T = 0$ K$)$ already in the valence bands, further lowering $\mu(T)$ at finite temperatures (still would happen due to the intrinsic broadening of the Dirac band) no longer produces topology change at $\mu(T)$ (Fermi pockets always maintain hole-like). This washes out the sign changing feature in *S* and leads to a monotonic $\rho_{xx}(T)$ profile (blue curves). In addition, with increasing *n*-type doping level, $T_p$ shifts to higher temperatures (following curves with black, cyan [30], magenta [33], and dark yellow [32] colors). The power law dependence $T_p \propto n_H^{0.27}$ [figure 5(c)], resembles a similarity to the characteristic $n_H^{1/3} \sim \varepsilon_F$, as dictated by the linear energy ($\varepsilon$) *vs.* momentum (*k*) dispersion in a 3D Dirac band [$\varepsilon(\mathbf{k}) = \hbar v_F k$], which further implies the dominance of the Dirac band at low temperatures. Furthermore, by lowering $\varepsilon_F$ via annealing (where excessive Te, serving as an *n*-type dopant in ZrTe$_5$, is reduced), we most recently achieved an even lower $T_p \sim 40 - 50$ K. We now identify that $T_p$ monotonically increases with $\varepsilon_F$, which can be used as a powerful experimental knob to control the Lifshitz transition, a feature rarely accessible in other Lifshitz systems [62, 63].

## 4. Conclusion

We report detailed transport measurements of single crystal ZrTe$_5$. The data are analyzed using the two-band model. The derived carrier concentration is in the order of $10^{16}$ cm$^{-3}$. The mobility in the Dirac band is extremely high with a value exceeding $3 \times 10^5$ cm$^2$ V$^{-1}$ s$^{-1}$. A temperature-induced Lifshitz transition is identified in the highly mobile Dirac band that supports dissipationless chiral charge transport under $\mathbf{E} \cdot \mathbf{B} \neq 0$. As temperature increases, the chemical potential moves from the upper Dirac band to the lower one. A secondary band hosting less-mobile carriers is found to contribute to the transport properties as well. At temperature near zero, the secondary conduction band near the band edge is populated by electrons. At elevated temperatures, the chemical potential drops into the secondary band gap. The contribution to transport from the secondary bands are from the thermally excited carriers, changing from the dominant *n*-type to the *p*-type. The Lifshitz transition in the Dirac band is believed to be origin of the resistivity "anomaly" in ZrTe$_5$. The "anomaly" peak temperature $T_p$ is controlled by the Fermi level $\varepsilon_F$ ($T_p$ monotonically increases with $\varepsilon_F$), that is an effective way to fine tune the transport properties of transition metal pentatellurides.

## 5. Acknowledgments

This work was supported by the U.S. Department of Energy, Office of Basic Energy Science, Materials Sciences and Engineering Division, under Contract No. DE-SC00112704.

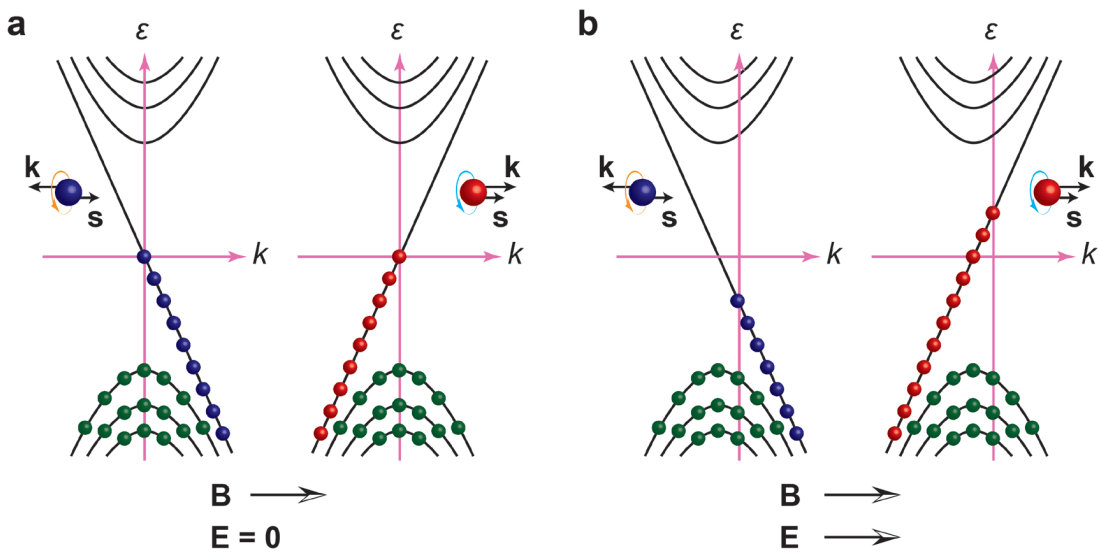

**Figure 1 | Chiral magnetic effect in Dirac semimetals.** (**a**) The left- and right-handed fermions occupying various Landau levels (LLs) in the presence of magnetic field **B**. On the lowest LL, the spins of positive (negative) chiral fermions are parallel (anti-parallel) to **B**. Therefore, for a positive fermion to be right-handed (i.e., have a positive projection of spin on momentum) means moving along the magnetic field, and for a negative fermion – moving against **B**. The left- and right-handed fermions are equally numbered under zero electric field **E**. (**b**) with **E** ∥ **B**, the positive (negative) fermions accelerate (decelerate) along **E** that is also parallel to **B**. This creates a non-zero chemical potential, leading to a net chiral magnetic effect current.

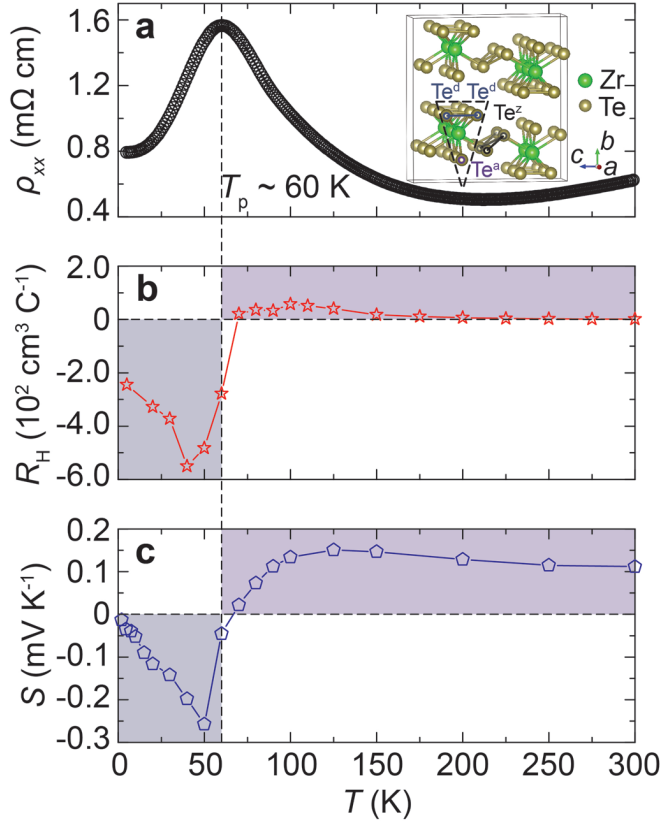

**Figure 2 | Basic properties of bulk ZrTe₅.** The temperature-dependent (**a**) electrical resistivity $\rho_{xx}$, (**b**) Hall coefficient $R_H$, and (**c**) Seebeck coefficient $S$. The dashed line, at the $\rho_{xx}$ peak temperature $T_p$ (approximately 60 K), indicates the resistivity "anomaly" occurs essentially at the temperature when both $R_H$ and $S$ change sign. The crystal structure of ZrTe₅ is illustrated in the inset of (**a**), highlighting the symmetrically non-equivalent apical (Te$^a$), dimer (Te$^d$), and zigzag (Te$^z$) tellurium atoms, along with the ZrTe₃ (dashed-line triangle) chains and the ZrTe₅ sheets.

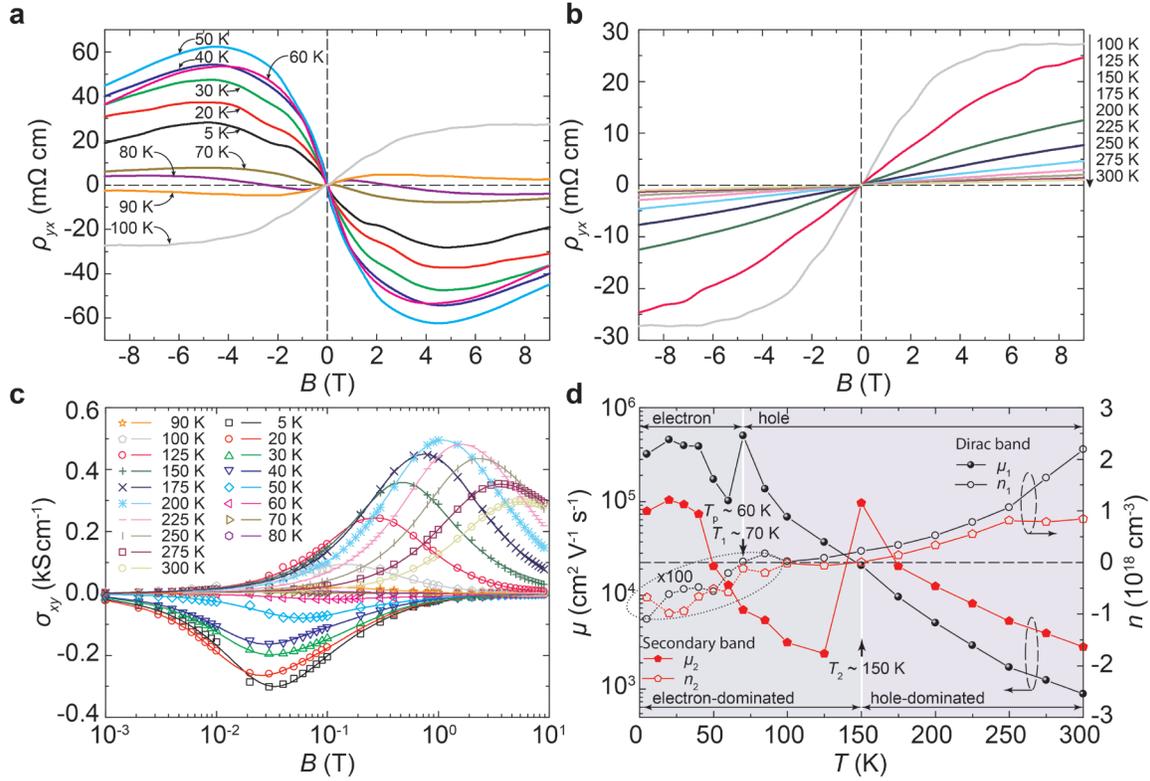

**Figure 3 | Multiple-carrier transport in bulk ZrTe$_5$.** The field dependence of Hall resistivity $\rho_{yx}(B)$ at (**a**) 5 – 100 K and (**b**) 100 – 300 K, respectively. (**c**) The calculated Hall conductivity $\sigma_{xy}$ (symbols) along with the two-band model fitting results (lines), showing a good agreement over all the measured temperature and magnetic field range. (**d**) The temperature dependence of the fitting parameters, namely, carrier concentration $n_1$ ($n_2$) and mobility $\mu_1$ ($\mu_2$), confirming contributions from the highly-mobile Dirac band and the less-mobile secondary band in transport processes. For better visibility, the data of $n_1$ ($n_2$) below 100 K (open symbols within the dotted ellipse, connected by dotted lines) are multiplied by a factor of 100. $T_1$ ($T_2$) identifies the temperature where the carrier changes type, i.e., from electrons to holes upon increasing temperature, in the Dirac (secondary) band.

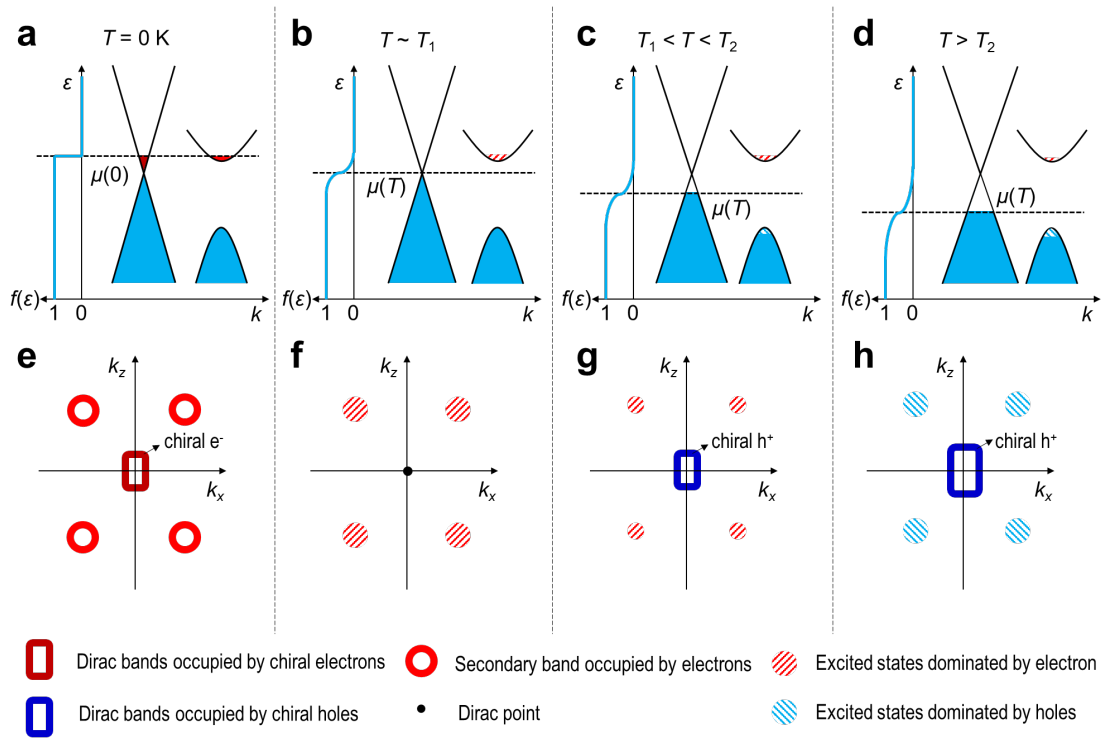

**Figure 4 | Lifshitz transition in bulk ZrTe₅.** (**a-d**) Evolution of the Dirac band at zone center and the secondary band, together with the Fermi-distribution function $f(\varepsilon)$ as a function of temperature. Accompany to the reduction of the Fermi velocity $v_F$ of the chiral quasiparticles is the broadening of the Dirac band as temperature increases. This increases the available states in the lower band that effectively lowers the chemical potential $\mu(T)$. (**e-h**) Sketches of the expected band topology at $\mu(T)$. For the Dirac band, the Fermi pocket changes from electron to hole at $T_1$, the rectangular shape approximately reflects the band anisotropy. The four off-centered circles at zero temperature are the electron pockets of the secondary band. At finite temperatures, in the secondary bands, there are thermally excited states (hatched areas). The dominant carriers in the secondary bands change from electrons to holes near $T_2$ as temperature increases. For clarity, the excitation in the Dirac band is not shown.

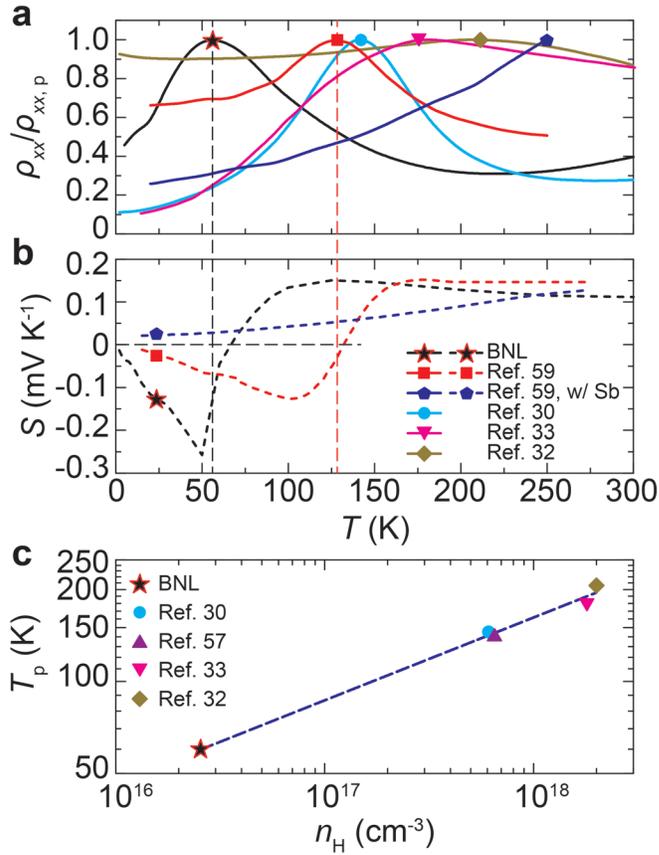

**Figure 5 | Electronic origin of the resistivity "anomaly" in bulk ZrTe$_5$.** Literature results of (**a**) electrical resistivity $\rho_{xx}$ and (**b**) Seebeck coefficient $S$ are compared with data from this work. The resistivity peak position $T_p$ always seems to coincide with the temperature where the sign of $S$ changes, regardless of significant differences in the background carrier concentration $n_H$ [this work (black) vs. Ref. [59] (red)]. On one hand, the Lifshitz transition can be effectively switched off, by setting the Fermi level $\varepsilon_F$ [i.e., $\mu(T = 0 \text{ K})$] deep into the valence bands (p-type doping via Sb [59]), where the downward shift of $\mu(T)$ (upon rising temperature) does not alter the band topology anymore (always p-type pockets). Hence $S$ is positive at all temperatures and the peak feature of $\rho_{xx}$ is completely washed out (blue). On the other hand, $T_p$ shifts towards higher temperatures upon increasing n-type doping, in the order of black, to cyan [30], magenta [33], and dark yellow [32] colors. (**c**) The log-log plot of $T_p$ vs. $n_H$, where the linear fit leads to approximately $T_p \propto n_H^{0.27}$. The power law dependence (rather close to the characteristic $n_H^{1/3} \sim \varepsilon_F$ for a linearly dispersed Dirac band) correlates $T_p$ directly with $\varepsilon_F$, suggesting a Dirac-band-dominated transport in bulk ZrTe$_5$ at cryogenic temperatures.

# SUPPLEMENTARY INFORMATION

**Lifshitz transition mediated electronic transport anomaly in bulk ZrTe$_5$**

Hang Chi[1†], Cheng Zhang[1,2†], Genda Gu[1], Dmitri E. Kharzeev[3,4], Xi Dai[5,6] and Qiang Li[1*]

[1]*Condensed Matter Physics and Materials Science Department,*
*Brookhaven National Laboratory, Upton, New York 11973, USA*
[2]*Materials Science and Engineering Department, Stony Brook University, Stony Brook, NY 11794, USA*
[3]*Department of Physics, Brookhaven National Laboratory, Upton, New York 11973, USA*
[4]*Department of Physics and Astronomy, Stony Brook University, New York 11794, USA*
[5]*Beijing National Laboratory for Condensed Matter Physics,*
*Institute of Physics, Chinese Academy of Sciences, Beijing 100190, China*
[6]*Collaborative Innovation Center of Quantum Matter, Beijing 100871, China*

[†]These authors contributed equally to this work.

[*]Correspondence should be addressed to Q.L. (email: qiangli@bnl.gov).

**Keywords:** ZrTe$_5$, Dirac semimetal, Lifshitz transition, and Chiral magnetic effect.

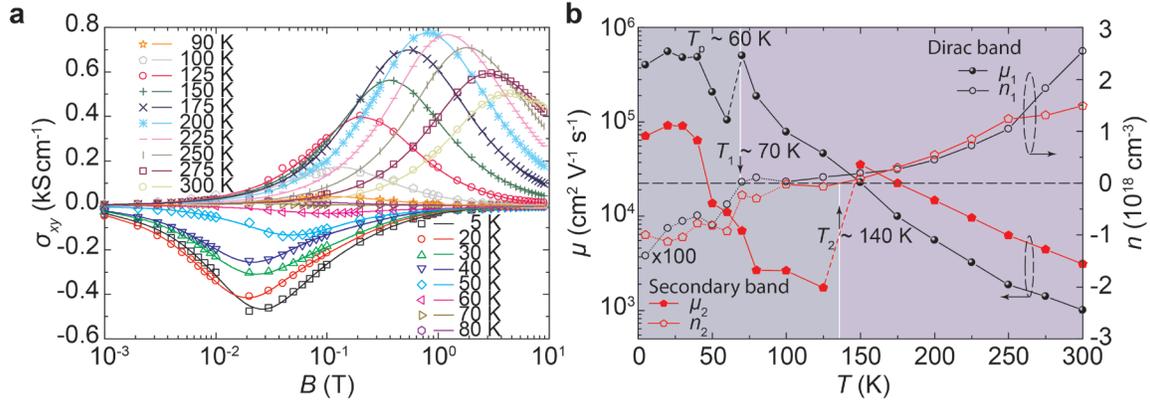

**Supplementary Figure 1 | Hall conductivity of bulk ZrTe$_5$.** (**a**) The Hall conductivity calculated via $\sigma_{xy}$ $[\equiv \rho_{yx}/(\rho_{xx}^2 + \rho_{yx}^2)]$, ignoring the in-plane anisotropy. (**b**) The derived fitting parameters. A similar diagram, assuming isotropic in-plane transport behavior, has also been obtained for HfTe$_5$ [1].

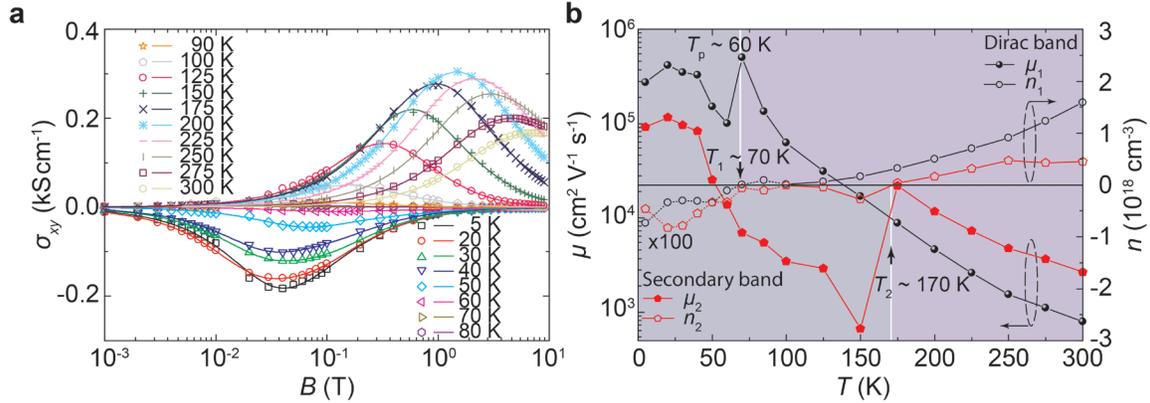

**Supplementary Figure 2 | Two-band modelling of bulk ZrTe$_5$.** (**a**) The Hall conductivity calculated via $\sigma_{xy}$ $[\equiv \rho_{yx}/(\rho_{xx}\rho_{yy} + \rho_{yx}^2)]$, assuming the in-plane (*a-c* plane) anisotropy $b = \rho_{yy}/\rho_{xx} = 4$. (**b**) The fitting parameters from this set of data reveal essentially the same physics regarding the Lifshitz transitions ($T_2$ is slightly different than that in the main text and/or in supplementary figure 1), indicating the numerical modelling is insensitive of the choice of *b*.

**Supplementary Table 1. Transport coefficients of ZrTe$_5$.**

| Reference | Thickness $t$ | Carrier concentration $n_\text{H}$ (cm$^{-3}$) | Mobility $\mu_\text{H}$ (10$^4$ cm$^2$ V$^{-1}$ s$^{-1}$) | Resistivity peak $T_\text{p}$ (K) |
|---|---|---|---|---|
| This work | 0.1 mm | $-2.6\times10^{16}$ | 30 | 60 |
| McGuire et al. [2]* | - | - | - | 130 |
| Zheng et al. [3] | 160 nm | $-6.1\times10^{17}$ | 4.6 | 145 |
| Izumi et al. [4] | - | $-6.5\times10^{17}$ | 7.9 | 140 |
| Yu et al. [5] | 620 nm | $-1.8\times10^{18}$ | 3 | 180 |
| Niu et al. [6] | 25 nm | $\sim -2\times10^{18}$ | - | 205 |
| Pariari et al. [7] | - | $+8.3\times10^{15}$ | - | - |

*The value of $n_\text{H}$ is not available in the original paper. But it mostly likely has much higher $n_\text{H}$ than our samples used in the current manuscript, as all the other iodine-grown samples do.

**Supplementary Note 1: Crystal structure of ZrTe$_5$.**

As shown in the inset of figure 1(a) in the main text, distinct types of Te are designated [8] as apical (Te$^a$), dimer (Te$^d$, bond distance $d_{\text{Te,d-Te,d}}$ = 2.717 Å), and zigzag (Te$^z$, $d_{\text{Te,z-Te,z}}$ = 2.946 Å), respectively. Zr, Te$^a$ and Te$^d$ form quasi-one-dimensional (1D) prismatic ZrTe$_3$ (dashed-line triangle) chains along the *a*-axis. The distance between Te$^d$-Te$^d$ dimers, in the chain direction, is the lattice constant $a$ = 3.9875 Å (while $b$ = 14.530 Å and $c$ = 13.724 Å). These 1D chains are held together along the *c*-axis via Te$^z$ zigzag chains, forming quasi-two-dimensional (2D) ZrTe$_5$ sheets (monolayers). The bonding amongst ZrTe$_3$ chains inside the ZrTe$_5$ sheets is likely of the ionic/covalent nature (weaker than that within the chain though), considering the distance between Zr and Te$^z$ ($d_{\text{Zr-Te,z}}$ = 2.965 Å) is on par with that constituting the ZrTe$_3$ prisms ($d_{\text{Zr-Te,a}}$ = 2.986 Å and $d_{\text{Zr-Te,d}}$ = 2.951 Å). The ZrTe$_5$ sheets are interconnected by weak van der Waals (vdW) forces (spacing $\Delta_\text{vdW}$ = $b$/2 = 7.265 Å) along the *b*-axis, resulting in a three-dimensional (3D) bulk crystal. The weak interactions in ZrTe$_5$ (needle-like crystals, grown preferably along the *a*-axis) warrant easy cleavage that produces 2D flakes (peeling off layers along the *b*-axis) and/or 1D fibers (breaking bonds in the *c*-axis).

**Supplementary References**